# Exact Maxwell evolution equation of resonators dynamics: temporal coupled-mode theory revisited

TONG WU,[1] PHILIPPE LALANNE[1,*]

[1]*Laboratoire Photonique, Numérique et Nanosciences (LP2N), IOGS- Université de Bordeaux-CNRS, 33400 Talence cedex, France*
*\*philippe.lalanne@institutoptique.fr*

**Abstract:** Despite its widespread significance, the temporal coupled-mode theory (CMT) lacks a foundational validation based on electromagnetic principles and stands as a phenomenological theory relying on fitted coupling coefficients. We employ an ab initio Maxwellian approach using quasinormal-mode theory to derive an "exact" Maxwell evolution (EME) equation for resonator dynamics. While the resulting differential equation bears resemblance to the classical one, it introduces novel terms embodying distinct physics, suggesting that the CMT predictions could be faulted by dedicated experiments, for instance carried out with short and off-resonance pulses, or with resonators of sizes comparable to or greater than the wavelength. Nonetheless, our examination indicates that, despite its inherent lack of strictness, the CMT enables precise predictions for numerous experiments due to the flexibility provided by the fitted coupling coefficients. The new EME equation is anticipated to be applicable to all electromagnetic resonator geometries, and the theoretical approach we have taken can be extended to other wave physics.

## 1. Introduction

The temporal coupled-mode theory (CMT) is an acclaimed and widely used theoretical framework for modeling the continuous wave response and temporal dynamics of any integrated or free-space photonic resonant structure. Initially developed for the analysis and design of resonant systems, as well as the comprehension of energy coupling into and out of cavities and its exchange among various resonant modes [1-4], CMT has evolved over three decades to encompass advanced systems and contemporary physical phenomena. These include noninstantaneous nonlinearities, gain, PT-symmetry, 2D photonic materials, and time-reversal, underscoring its vast capabilities.

CMT holds the potential to simplify the description of resonant systems with arbitrary material compositions and geometric configurations. Even intricate systems can be treated as lumped harmonic oscillators. At its core, the dynamics of the resonator can be reduced to an ordinary differential equation with respect to time

$$\frac{da_m}{dt} = -i\widetilde{\omega}_m a_m + \langle \kappa_m^* | s_+ \rangle, \quad (1)$$

where $|a_m|^2$ represents the energy stored in the $m^{th}$ mode of the resonator and $\widetilde{\omega}_m$ is the complex-valued frequency of the mode, encompassing radiative decay to excitation channels, free space and absorption. We intentionally use conventional notations in Eq. (1) and denote by $|s_+\rangle$ the vector of input amplitudes and $|\kappa_m\rangle$ the vector of coupling constants between the mode and the channels.

The evolution equation (1) is typically accompanied by an additional linear equation that delineates the decay of the resonator into each individual channel. Generally, the coupling coefficients introduced by these additional equations are derived with multiple smart assumptions. We will not delve into them here, maintaining our focus on the fundamental dynamics of the resonator.

Equation (1) stands as a pivotal outcome of CMT, offering high intuitiveness, computational simplicity, and a clear delineation of the physics governing resonances. However, it should be regarded as a heuristic approach, lacking theoretical support for its validity. In standard usage, physical quantities characterizing the cavity (e.g., resonance frequency and coupling constants) are fitted, and their extraction is generally believed to be rigorously feasible through numerical or experimental means.

The primary motivation behind this work was to establish a rigorous foundation for Eq. (1) and derive an analytical expression for it directly from Maxwell equations. In this pursuit, we discovered that the evolution equation is only an approximation. Our objectives are thus to elucidate the approximations inherent in the CMT model, comprehend their consequences, and raise awareness when the differences become consequential. We contend that the approximations are significant enough to warrant a dedicated publication, building upon an analysis initially presented in a prior publication [5] by the same group.

Except for the study in [6], as far as we are aware, there is a lack of any Maxwellian derivation of the CMT in the existing literature. Furthermore, based on our comprehension of the analysis in [6], it appears that the Maxwellian evolution equation derived in our current work differs from that derived in [6]. Also, the accuracy of the CMT evolution equation under specific conditions varies.

In Section 2, we initiate our exploration from Maxwell equations, undertaking an ab initio derivation of the differential equation governing the dynamics of resonators. This derivation leads us to an equation distinct from Eq. (1), which we shall refer to as the "exact" Maxwell evolution (EME) equation. The ability to derive this is facilitated by the advancements achieved over the past decade in electromagnetic quasinormal modes (QNMs) [7].

In Section 3, we highlight the two primary distinctions between the CMT evolution equation and the EME equation: the coupling coefficient is non-local, and the excitation is not directly proportional to the driving field, but to its temporal derivative. We also conduct a comprehensive assessment through a generic example, enabling us to discern the circumstances under which the predictions of the CMT and EME equations significantly differ.

In section 4, we quantitatively test the EME equation for a real optical system. With this example, we show that a term related to the temporal derivative of the driving field plays a significant role in determining the excitation of QNMs, and this term is not given by the CMT equation.

The insights gleaned from our conclusions in the realm of electromagnetism hold the potential to extend their influence on other types of waves for which QNM theory offers robust support [8].

## 2. The exact Maxwell evolution equation

The EME equation is derived by employing electromagnetic quasinormal mode (QNM) theory, which has experienced substantial advancements over the past decade and has now reached a level of maturity that allows its secure application to numerous contemporary electromagnetic challenges in photonics and plasmonics. Electromagnetic QNMs are source-free solutions of Maxwell equations, $\nabla \times \tilde{\mathbf{E}}_m = -i\tilde{\omega}_m \mu_0 \tilde{\mathbf{H}}_m$, $\nabla \times \tilde{\mathbf{H}}_m = i\tilde{\omega}_m \boldsymbol{\varepsilon}(\tilde{\omega}) \tilde{\mathbf{E}}_m$, which satisfy the outgoing-wave condition for $|\mathbf{r}| \to \infty$. $\tilde{\mathbf{E}}_m$ and $\tilde{\mathbf{H}}_m$ respectively denote the *normalized* electric and magnetic fields [7], $\boldsymbol{\varepsilon}$ denotes the possibly-dispersive permittivity tensors. The QNM field exponentially decays in time and $\text{Im}(\tilde{\omega}_m) < 0$ with the $\exp(-i\omega t)$ notation.

Hereafter, we consider non-magnetic materials, with $\mu_0$ denoting the permeability of vacuum. Extending the analysis to magnetic cases is straightforward [7]. We also assume frequency-independent permittivity for simplicity, emphasizing that the distinctions we uncover between the evolution equations are unrelated to dispersion. The EME equation for resonators composed of Drude-Lorentz materials will be presented after the derivation of the EME equation. We will not derive it since it was previously established in [5].

Subsequently, we adopt the QNM theory with a complex-mapping regularization. This approach has produced highly compelling results, particularly in addressing intricate electromagnetic problems such as those involving branch cuts due to the presence of substrates [5]. The concept of QNM regularization with complex mapping originated in quantum mechanics in the late 1970s. In the realm of electromagnetism, the favored mapping employs a complex scaling transformation known as perfectly matched layers (PMLs). This transformation maps the inherently divergent QNMs of open space onto square-integrable regularized QNMs [9]. For an in-depth examination of QNM regularization, including the use of complex mapping, refer to Section 5.3 in [10].

The regularized QNMs exhibit numerous similarities with the normal modes of Hermitian systems. Specifically, they are square-integrable, can undergo normalization, and can be employed to expand the electromagnetic field $\mathbf{\Psi}_s(\mathbf{r},\omega)\exp(-i\omega t)$ scattered by a resonator illuminated by a monochromatic field oscillating at frequency $\omega$

$$\mathbf{\Psi}_s(\mathbf{r},\omega) = \sum_m \alpha_m(\omega)\tilde{\mathbf{\Psi}}_m(\mathbf{r}), \qquad (2)$$

where $\alpha_m$ is the complex-valued excitation coefficient of the $m^{th}$ QNM. Throughout the manuscript, $\mathbf{\Psi}$ denotes a bivector formed by electric and magnetic fields, for instance $\tilde{\mathbf{\Psi}}_m = [\tilde{\mathbf{E}}_m, \tilde{\mathbf{H}}_m]$.

The extension of Eq. (2) is complete for $\mathbf{r}$ in the interior of compact resonators placed in free space [10,11]. We will consider this special case (a simple 1D slab in air) in the final numerical test. For non-uniform backgrounds, e.g. layered substrate, there are branch cuts. To account for this possibility, we assume that the extension of Eq. (2) encompasses not only QNMs that are unaffected by the PML mapping but also an infinite set of "numerical" eigenvectors. These eigenvectors are computed as source-free solutions of Maxwell equations within the mapped space. Utilizing this expanded basis, highly accurate reconstructions of the field scattered by the resonator have been achieved, even for complex geometries such as resonators with dispersive materials positioned on substrates. Accurate reconstructions have been reported in both spectral [10,12-17] or temporal [5,12] domains.

The QNM framework based a complex-mapping regularization provide a unique expression for the excitation coefficient [5,13]

$$\alpha_m(\omega) = \frac{\omega}{\tilde{\omega}_m - \omega} \int_{V_r} \Delta\boldsymbol{\varepsilon}(\mathbf{r})\mathbf{E}_b(\mathbf{r},\omega) \cdot \tilde{\mathbf{E}}_m d^3\mathbf{r}, \qquad (3)$$

for the case of nondispersive materials under consideration here.

This expression holds significant importance as the subsequent mathematical analysis depends on it. It's worth noting that all existing QNM frameworks, not just the one employed here, suggest an identical expression for non-dispersive materials, as indicated in Table 1 of reference [7]. For dispersive materials – a case that will be addressed at the end of the Section – several expressions exist, see Table 1 in reference [7] and the comprehensive discussion in reference [16].

Except for the pole prefactor, $\alpha_m$ essentially represents an overlap integral between the QNM (or numerical modes) and the background field $\mathbf{\Psi}_b = [\mathbf{E}_b, \mathbf{H}_b]$ in the absence of resonator. $\Delta\boldsymbol{\varepsilon}(\mathbf{r})$ specifies the permittivity variation used for the scattered-field formulation, given by $\Delta\boldsymbol{\varepsilon}(\mathbf{r}) = \boldsymbol{\varepsilon}_R(\mathbf{r}) - \boldsymbol{\varepsilon}_b(\mathbf{r})$, where $\boldsymbol{\varepsilon}_R(\mathbf{r})$ and $\boldsymbol{\varepsilon}_b(\mathbf{r})$ denote the permittivities of the resonator and the background, respectively. $\Delta\boldsymbol{\varepsilon}(\mathbf{r}) \neq 0$ within a compact volume $V_r$ that defines the resonator in the scattered-field formulation [7].

Unlike other rigorous methods featuring a finite number of QNMs and a background term [18,19], the intrinsic strength of the regularization approach for theoretical works – like the present one – lies in providing a unified mathematical framework. Within this framework, all eigenstates, QNMs or numerical modes, are treated identically. Owing to the analytical $\omega$-dependence of the excitation coefficient, the resonator response to any incident driving wavepacket can be analytically calculated using a spectral decomposition, summing over all distinct frequencies. Subsequently, the field $\boldsymbol{\Psi}_S(\boldsymbol{r},t)$ scattered by the resonator illuminated by a wavepacket can be expressed as a sum of contributions from QNMs and numerical modes [7]

$$\boldsymbol{\Psi}_S(\boldsymbol{r},t) = \mathrm{Re}\big(\sum_m \beta_m(t)\widetilde{\boldsymbol{\Psi}}_m(\boldsymbol{r})\big), \tag{4}$$

with $\beta_m(t) = \int_{-\infty}^{\infty} \alpha_m(\omega) \exp(-i\omega t)\, d\omega$.

Using a spectral decomposition of the incident wavepacket, $\mathbf{E}_b(\boldsymbol{r},\omega) = (2\pi)^{-1}\int_{-\infty}^{\infty} \mathbf{E}_b(\boldsymbol{r},t)\exp(i\omega t)\,dt$, and replacing $\alpha_m(\omega)$ with Eq. (3), we get

$$\beta_m(t) = \int_{-\infty}^{\infty}\left(\int_{-\infty}^{\infty}\frac{\omega}{2\pi(\widetilde{\omega}_m-\omega)}O_m(t')\exp\big(i\omega(t'-t)\big)dt'\right)d\omega. \tag{5}$$

Here the overlap is now defined as $O_m(t) = \int_{V_r}\Delta\boldsymbol{\varepsilon}(\boldsymbol{r})\mathbf{E}_b(\boldsymbol{r},t)\cdot\widetilde{\mathbf{E}}_m d^3\boldsymbol{r}$.

Hereafter, the EME equation is derived from Eq. (5) utilizing complex analysis and generalized functions – distributions – with a particular emphasis on Cauchy's integral theorem. Another demonstration, which is straightforward and explicit, is currently under elaboration [20].

For dispersive PMLs characterized by a complex scaling factor inversely proportional to the frequency, causality is preserved, and all eigenfrequencies of the regularized problem are situated in the lower half of the complex plane. As a result, all temporal excitation coefficients $\beta_m(t)$ can be analytically computed utilizing contour integration and the Cauchy integral formula for the pole $\widetilde{\omega}_m$.

We recast the temporal integral in Eq. (5) in three terms,

$$\beta_m(t) = \int_{-\infty}^{t-\epsilon}\int_{-\infty}^{\infty}\ldots d\omega dt' + \int_{t-\epsilon}^{t+\epsilon}\int_{-\infty}^{\infty}\ldots d\omega\, dt' + \int_{t+\epsilon}^{\infty}\int_{-\infty}^{\infty}\ldots d\omega\, dt' = A + B + C, \tag{6}$$

where the three integrands are strictly identical to that in Eq. (5) and are represented by '…' hereafter. $\epsilon$ is a small and positive quantity, and we will consider the limit $\epsilon \to 0$ later on. As $\omega \to \infty$, the exponential factor $\exp\big(i\omega(t'-t)\big)$ has different behaviors for these three terms. We will show later that $A$, $B$, and $C$ can be computed analytically by wisely selecting the contour integration paths shown in Fig. 1.

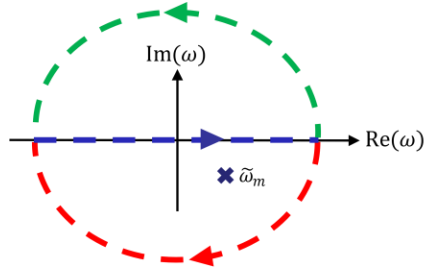

**Fig. 1.** The two trajectories (red and green arcs) of the complex frequency plane used in the derivation of the EME equation.

Let us first consider the last term, $C$. To compute the integral, we utilize the upper-plane green arc. Since $\exp\big(i\omega(t'-t)\big)$ vanishes as $\mathrm{Im}(\omega) \to \infty$ and there is no pole in the upper plane, we conclude that $C$ equals 0.

Now, let us focus on term $A$. In this case, the integral is null on the red arc, and the frequency integral can be replaced by the residue, $2\pi i\widetilde{\omega}_m O_m(t')\exp\big(i\widetilde{\omega}_m(t'-t)\big)$. Taking the limit $\epsilon \to 0$, we find $A = i\widetilde{\omega}_m\int_{-\infty}^{t}O_m(t')\exp\big(i\widetilde{\omega}_m(t'-t)\big)dt'$.

For term $B$, we use the equality $\frac{\omega}{(\widetilde{\omega}_m-\omega)} = \frac{\widetilde{\omega}_m}{(\widetilde{\omega}_m-\omega)} - 1$ and obtain $B = B_1 + B_2$ with $B_1 = \int_{t-\epsilon}^{t+\epsilon}\int_{-\infty}^{\infty}\frac{\widetilde{\omega}_m}{\omega}\ldots d\omega\, dt'$ and $B_2 = -\int_{t-\epsilon}^{t+\epsilon}\int_{-\infty}^{\infty}(2\pi)^{-1}O_m(t')\exp\big(i\omega(t'-t)\big)d\omega dt'$. Using the $\omega$-Fourier transformation of the delta function, one readily finds that $B_2 = -\int_{t-\epsilon}^{t+\epsilon}\delta(t'-$

$t)O_m(t')\,dt' = -O_m(t)$. To demonstrate that $B_1$ is null, we cast $B_1 = (2\pi)^{-1} \int_{t-\epsilon}^{t+\epsilon} \int_{-\infty}^{\infty} \frac{\widetilde{\omega}_m}{(\widetilde{\omega}_m - \omega)} O_m(t') \exp(i\omega(t'-t)) d\omega dt' = \int_{t-\epsilon}^{t+\epsilon} \int_{-\infty}^{\infty} \frac{\widetilde{\omega}_m}{\omega} \ldots d\omega dt'$ into two parts, $B_1 = B_1' + B_1'' = \int_{t-\epsilon}^{t} \int_{-\infty}^{\infty} \frac{\widetilde{\omega}_m}{\omega} \ldots d\omega dt' + \int_{t}^{t+\epsilon} \int_{-\infty}^{\infty} \frac{\widetilde{\omega}_m}{\omega} \ldots d\omega dt'$. $B_1'$ can be calculated with the residue theorem with the red arc since $t' - t \leq 0$. The residue theorem gives us $B_1' = i\widetilde{\omega}_m \int_{t-\epsilon}^{t} O_m(t') \exp(i\widetilde{\omega}_m(t'-t))\,dt' + \int_{t-\epsilon}^{t} \int_{red\,arc} \frac{\widetilde{\omega}_m}{\omega} \ldots d\omega dt'$. By parametrizing the arc path with $\omega = a \exp(i\theta)$ with $a \to \infty$ and $\theta \in [0, -\pi]$, we show that the modulus of the arc integral is bounded by a finite positive number. Thus, taking the limit $\epsilon \to 0$ in the temporal integral, we obtain $B_1' = 0$. An identical same reasoning holds for $B_1''$ by considering the green arc. There is no pole contribution and the arc integral being bounded for the same reason, we have $B_1'' = 0$.

Altogether, we have

$$\beta_m(t) = i\widetilde{\omega}_m \int_{-\infty}^{t} O_m(t') \exp(i\widetilde{\omega}_m(t'-t)) dt' - O_m(t), \quad (7)$$

and upon applying the derivative with respect to $t$ on both sides of the equation, we obtain the EME equation

$$\frac{d\beta_m(t)}{dt} = -i\widetilde{\omega}_m \beta_m(t) - \frac{d}{dt} \langle \mathbf{E}_b(\mathbf{r},t) | \Delta \varepsilon(\mathbf{r}) | \widetilde{\mathbf{E}}_m(\mathbf{r}) \rangle, \quad (8)$$

in which, for convenience, we introduce the notation $\langle \mathbf{E}_b | X | \widetilde{\mathbf{E}}_m \rangle = \int_{V_r} X(\mathbf{r}) \mathbf{E}_b(\mathbf{r},t) \cdot \widetilde{\mathbf{E}}_m(\mathbf{r}) d^3\mathbf{r}$ with $X$ an arbitrary $3 \times 3$ matrix. We intend to publish another demonstration of Eq. (8) at a later time, which will not involve the combination of Schwartz distributions and complex analysis.

We recall that Eq. (8) holds for non-dispersive materials. A similar derivation can be conducted for dispersive materials [5] using the auxiliary-field method [21,22] and an analytical expression for the permittivity. For instance, for resonators with a standard multi-pole Lorentz-Drude permittivity and for a dispersionless background in the volume $V_r$ ($\boldsymbol{\varepsilon}_b(\mathbf{r})$ may depend on the frequency for $\mathbf{r} \notin V_r$), it was previously shown [5] that

$$\frac{d\beta_m(t)}{dt} = -i\widetilde{\omega}_m \beta_m(t) + i\widetilde{\omega}_m \langle \mathbf{E}_b | \boldsymbol{\varepsilon}(\mathbf{r}, \widetilde{\omega}_m) - \boldsymbol{\varepsilon}_\infty(\mathbf{r}) | \widetilde{\mathbf{E}}_m \rangle - \frac{d}{dt} \langle \mathbf{E}_b | \boldsymbol{\varepsilon}_\infty(\mathbf{r}) - \boldsymbol{\varepsilon}_b(\mathbf{r}) | \widetilde{\mathbf{E}}_m \rangle, \quad (9)$$

where $\varepsilon_\infty$ denotes the high-frequency permittivity constant of the resonator. Equation (9) has been quantitatively tested for a gold bowtie antenna illuminated by a Gaussian pulse with a central frequency in the visible range, see Fig. 2 in [5]. It will not be further tested hereafter.

Equations (8) or (9) share similarities with the CMT evolution equation, yet differences are apparent, underscoring the incorporation of heuristics in the CMT. These heuristics are discussed in the next section.

Further deeper insights into Maxwellian derivation of the CMT regarding outcoupling with ports can be found in reference [6]. However, it appears that Eqs. (8) or (9) deviate from Eq. (45) in reference [6], notably in terms of the final term, which incorporates the temporal derivative of the incident electric field $\mathbf{E}_b(\mathbf{r},t)$.

Additionally, the following qualitative discussion regarding instances where the CMT and EME evolution equations diverge differs from the one presented in reference [6].

## 3. Qualitative discussion

We observe three primary distinctions between the EME equation and the classical CMT evolution equation:

- In the CMT, the squared absolute value of the modal excitation coefficient, $|a_m|^2$, signifies the energy stored in the mode $m$ of the resonator. The energy stored in QNMs lacks meaning in a non-Hermitian context: the mode energy is not square integrable since the QNM fields exponentially diverge outside the resonator, see related discussions on the quality factor or mode volume in [7].
- The second difference, unrelated to hermiticity, arises from the CMT assumption that the mode excitation by the wavepacket occurs locally [3,4]. The local excitation approximation assumes that $s_+$ is proportional to the wavepacket electric field $\mathbf{E}_b(\mathbf{r}_c, t)$ at an arbitrary point $\mathbf{r}_c$ typically chosen inside the resonator volume. In contrast, the QNM approach predicts a nonlocal excitation that spans the entire resonator volume through a spatial convolution of the incident wavepacket and QNM fields. Only in scenarios where the resonator is deep subwavelength, e.g. lump element circuits, does $O_m(t)$ become directly proportional to the incident wavepacket.
- The third difference has not been predicted in prior works (except in [5]) and is unexpected: it highlights that the driving force is proportional to the derivative of the incident field. We do not precisely know the physical origin of this term. Nevertheless, we note that it implies that the excitation coefficient $\beta$ responds instantaneously to the driving field, as observed with the last term in Eq. (7). We further note that this term is absent in the advanced formulation of Eq. (9) when dispersive materials are considered, specifically for $\varepsilon_\infty = \varepsilon_b$. For most materials, the high-frequency dielectric constant is anticipated to be approximately 1, indicating that materials behaves as if they were transparent or non-polarizable at high frequency. We have verified that, with the inclusion of background permittivity dispersion, this term vanishes for $\varepsilon_R(\omega \to \infty) = \varepsilon_b(\omega \to \infty) \approx 1$. Nonetheless, it is essential to highlight that the derivative term should be considered in usual simplified models assuming non-dispersive dielectrics.

The temporal CMT has been extensively utilized for several decades, consistently demonstrating its relevance in modeling the temporal dynamics of both integrated and free-space photonic resonant structures. The whole literature does not report any substantial deviations between CMT predictions and experimental or full-numerical data. Hereafter, we try to understand why. Specifically, we elucidate why the CMT evolution Eq. (1) can yield predictions in very good agreement with those provided by the EME equation, despite the inherent mathematical disparities between the two formulations.

In this section, the analysis is carried out qualitatively with the aim of providing a broad discussion applicable to various systems, such as small plasmonic cavities or ring resonators. The next Section will provide a quantitative examination for a Fabry-Perot cavity.

**Slowly-varying envelop approximation.** The first approximation arises from the fact that incident wavepackets can often be described as the product of a slowly-varying envelope function, denoted as $H(t)$, and a fast carrier signal, $\exp(-i\omega t)$. Thus, the electric field of the incident wavepacket can be expressed as $\mathbf{E}_b(\mathbf{r}, t) = \mathbf{E}_0 H\left(t - \mathbf{r} \cdot \frac{\mathbf{u}}{c}\right) \exp\left[-i\omega\left(t - \mathbf{r} \cdot \frac{\mathbf{u}}{c}\right)\right]$, $\mathbf{u}$ being a unitary vector parallel to the incident wavevector. Under the slowly-varying envelop approximation, $\frac{d}{dt}\langle \mathbf{E}_b(\mathbf{r}, t) | \Delta \boldsymbol{\varepsilon}(\mathbf{r}) | \tilde{\mathbf{E}}_m \rangle$ in Eq. (8) approximately becomes $-i\omega \langle \mathbf{E}_b(\mathbf{r}, t) | \Delta \boldsymbol{\varepsilon}(\mathbf{r}) | \tilde{\mathbf{E}}_m \rangle$. In the next Section, we will show that this approximation is only valid for on-resonance pulses whose central frequency (or carrier frequency) matches the QNM frequency ($\omega \approx \tilde{\omega}_m$).

**Local excitation approximation.** Another commonly encountered scenario involves resonators characterized by subwavelength dimensions, such as RLC lump-element circuits. In these instances, it is reasonable to assume that the incident wavepacket does not vary at the scale of the resonator, and the overlap integrals $\langle \mathbf{E}_b | \Delta \boldsymbol{\varepsilon}(\mathbf{r}) | \tilde{\mathbf{E}}_m \rangle$ reduces to

$\langle \mathbf{E}_b(\mathbf{r}_c, t) | \Delta \varepsilon(\mathbf{r}_c) | \tilde{\mathbf{E}}_m \rangle$, $\mathbf{r}_c$ being the 'center of coupling' of the resonator. This approximation holds true even for significantly larger systems, including ring resonators or photonic crystal cavities, which are typically much larger than the wavelengths of the confined radiation. In such cases, the input port typically consists of a waveguide, and the overlap integral $\langle \mathbf{E}_b(\mathbf{r}_c, t) | \Delta \varepsilon(\mathbf{r}_c) | \tilde{\mathbf{E}}_m \rangle$ remains predominantly localized. This is primarily due to the fact that the evanescent tail of the waveguide mode generally only intersects a small portion of the cavity. Typically, the overlap extends over a volume no greater than a fraction of $\lambda^3$.

By utilizing these two approximations, we effectively achieve a primary goal of this study: deriving an analytical expression for the coupling constant outlined in Eq. (1)

$$\langle \kappa_m^* | s_+ \rangle \equiv i\omega \, \langle \tilde{\mathbf{E}}_m | \Delta \varepsilon(\mathbf{r}_c) | \mathbf{E}_b(\mathbf{r}_c, t) \rangle, \tag{10}$$

where $\tilde{\mathbf{E}}_m$ represents the electric field of the *normalized* QNM and $\omega$ is the carrier frequency.

The prevalence of slowly-varying envelopes and local excitations in many applications and the absence of a QNM framework in the recent past may explain why earlier experiments did not reveal any discrepancy in CMT predictions.

We now provide a qualitative assessment of the impact of these two approximations.

For the simulation, we consider a parallelepiped resonator characterized by an isotropic spatially uniform $\Delta \varepsilon$ (see Fig. 2). The resonator is assumed to be illuminated by a pulse with an electric field $\mathbf{E}_0 \, S(u) \exp(-i\omega u)$, where $u = t - (x - x_c)/c$, $c = 1$, $\omega = 1 - 0.05i$ is a complex frequency and $S(u) = (1 + e^{-\rho u})^{-1}$ is the sigmoid function, approaching a Heaviside function for large values of $\rho$. Note that the carrier frequency $\omega$ nearly matches the resonance frequency $\tilde{\omega}_m = 1.1 - 0.0025i$. The pulse is assumed to hint the 'center of coupling' $x_c$ of the resonator at time $t = 0$.

To simplify our qualitative analysis, we also assume that the electric field of the QNM is spatially uniform in the volume $V_r = As$. Thus, the overlap integral in Eq. (8) simplifies to $A \, \Delta \varepsilon \, \tilde{\mathbf{E}}_m \cdot \mathbf{E}_0 \, \langle S(u) \exp(-i\omega u) \rangle_x$. Here, the parameters $A$, $\Delta \varepsilon$, $\tilde{\mathbf{E}}_m$ and $\mathbf{E}_0$ are scaling parameters and their values are inconsequential for the comparison. What is crucial for assessing finite size effects is the resonator length, $s$, along the direction of propagation of the driving pulse (see Fig. 2).

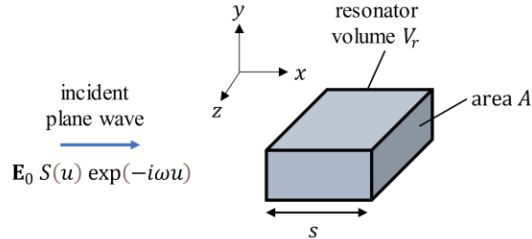

**Fig. 2.** Parameters used for the simulations reported in Fig. 3 for various pulse shapes ($\rho$) and resonator dimensions ($s$) characterized by the parameter $s$ along the direction $x$ of the incident plane wave $\mathbf{E}_0 \, S(u) \exp(-i\omega u)$.

Complex frequency illuminations have recently attracted significant interest for a range of intriguing applications, such as loss compensation [23]. In our investigation, we examine them to conveniently adjust the sharpness of the pulse envelope. Two values of $\rho$ are considered: $\rho = 0.1$ (gentle envelop, Fig. 3a) in and $\rho = 1$ (steep envelop, Figs. 3b and 3c).

Figure 3 summarizes the results of a comparison of the predictions obtained with the EME equation (cyan curve) and Eq. (10) (red curve), for increasing resonator sizes ($s$) and two pulse shapes ($\rho$) shown in grey.

We start with a subwavelength resonator ($s = 0.1\lambda$) and a gentle wavefront ($\rho = 0.1$) that resembles a Gaussian pulse. Consistently with other studies (not reported for compactness) with Gaussian pulses, we find a perfect agreement, see Fig. 3a.

To further test the predictive force of Eq. (10), we keep $s = 0.1\lambda$ and consider a steep envelop ($\rho = 1$). Again, we found a remarkable agreement with numerical data obtained directly by solving Eq. (8). This result is not reported in Fig. 3. Then we increase the resonator size ($s = 0.5\lambda$). We observe a discrepancy. However, if we multiply the prediction of Eq. (10) by 0.4, the rescaled data aligns perfectly with the prediction of the EME equation, see Fig. 3b. This is a noteworthy result. The CMT with a fitted coupling constant remains very accurate even when the resonator size is half a wavelength and fast-varying envelopes.

We finally consider a resonator size of one wavelength ($s = 1\lambda$). In this case, even with a large rescaling parameter, the CMT cannot predict the double peak response obtained with the EME equation (Fig. 3c). This implies that, even with fitted coupling coefficients, the CMT equation cannot be accurate and the EME formula emerges as especially important for investigating the fast dynamics of large-sized optical cavities involving linear or nonlinear effects [24].

**Illumination by pure harmonic waves.** Before we draw our conclusions, let us mention the special case where the resonator is illuminated by a harmonic oscillation wave, $\mathbf{E}_b = \mathbf{E}_0 \exp\left[-i\omega\left(t - \mathbf{r}\cdot\frac{\mathbf{u}}{c}\right)\right]$. It follows that $\frac{d}{dt}\langle \mathbf{E}_b(\mathbf{r},t)|\Delta\varepsilon(\mathbf{r})|\tilde{\mathbf{E}}_m\rangle$ in Eq. (8) transforms into $-i\omega \exp(-i\omega t)\langle \mathbf{E}_0 \exp\left(i\omega\mathbf{r}\cdot\frac{\mathbf{u}}{c}\right)|\Delta\varepsilon(\mathbf{r})|\tilde{\mathbf{E}}_m\rangle$. By introducing the driving signal at the coupling center, we derive a closed-form expression for $\kappa_m$

$$\kappa_m = i\omega\,|\mathbf{E}_0|^{-1}\exp\left[-i\omega\mathbf{r}_c\cdot\frac{\mathbf{u}}{c}\right]\langle \mathbf{E}_b(\mathbf{r},0)|\Delta\varepsilon(\mathbf{r})|\tilde{\mathbf{E}}_m\rangle. \tag{11}$$

This expression remains accurate regardless of the size of the resonator. It explains why the CMT has been successfully employed with monochromatic illumination for large resonators, such as racetrack type resonators, for which the coupling sections may span several tenths of wavelengths.

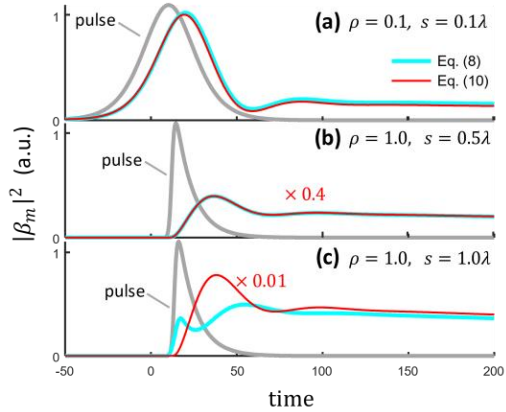

**Fig. 3.** Comparative analysis of the predictions obtained with the EME equation (8) (cyan curves) and the CMT evolution equation (1) with the analytical coupling constant given by Eq. (10) (red curve). We consider several resonator sizes ($s = 0.1, 0.5$ and $1\lambda$) and pulse shapes ($\rho = 0.1$ and 1) depicted with grey curves for clarity reasons. Panels **(a)**, **(b)**, and **(c)** depict scenarios with gentle sigmoid and deep subwavelength resonator size (high accuracy for Eq. (10)), steep sigmoid and subwavelength size (limited accuracy for Eq. (10)), and steep sigmoid and wavelength-scale resonator (significant inaccuracy and failure for Eq. (10)). Key parameters are $\omega = 1 - 0.05i$, $\tilde{\omega}_m = 1.1 - 0.0025i$. The resonator sizes are given in units of the wavelength ($\lambda = 2\pi c/\text{Re}(\omega)$). Note that the red curves in panels (b) and (c) are rescaled by 0.4 and 0.01, respectively.

Despite the heuristic nature of its derivation, the classical CMT Eq. (1), with a fitted coupling constant, proves to be robust for most cases. This robustness arises from a fundamental mathematical property of the exponential function: its preservation under differentiation or integration. This property approximately implements a proportional relationship between the carrier signal and the derivative term and the nonlocal overlap of the EME equation.

## 4. Numerical results

In the previous section, we have provided a qualitative discussion of under which conditions the CMT evolution Eq. (1) can yield accurate predictions. In this section, we quantitatively test the EME Eq. (8) for a real optical system. Our aim is to demonstrate the significance of the unexpected temporal-derivative term.

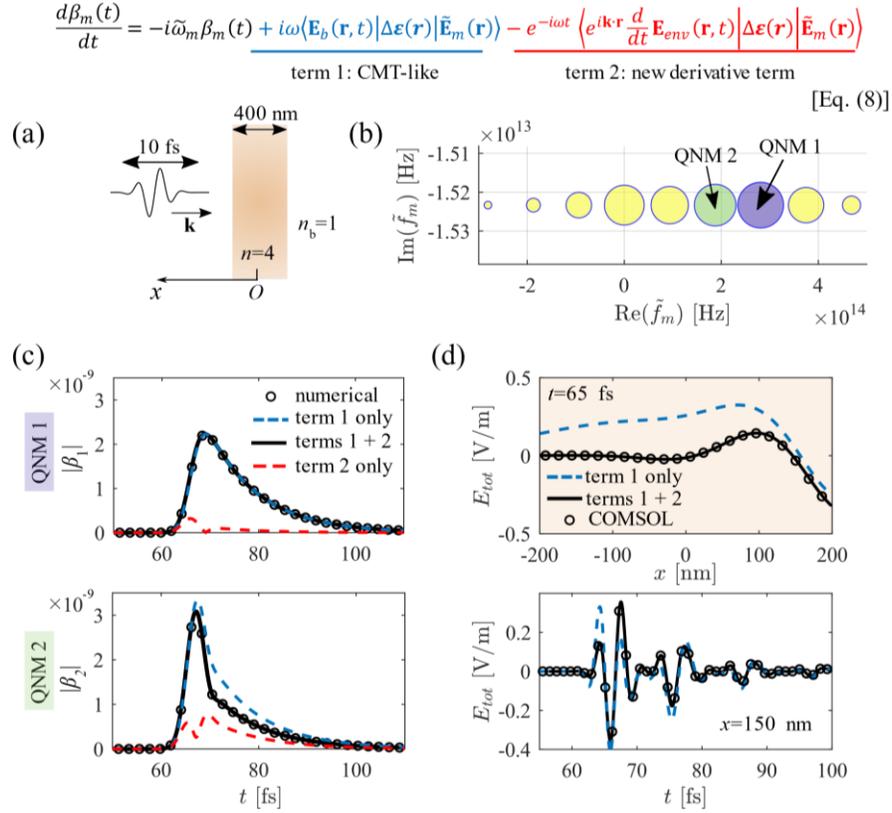

**Fig. 4.** Quantitative test of the EME Eq. (8) for a 1D Fabry-Perot resonator illuminated by a short pulse. **(a)** A 10-fs Gaussian pulse is incident upon a Fabry-Perot resonator of thickness 400 nm and refractive index $n = 4$. **(b)** QNMs of the resonator in the complex frequency plane ($\tilde{\omega}_m = 2\pi \tilde{f}_m$). The size of the markers visualizes the spectrally averaged $|\beta_m|$ obtained from Eq. (7). The real part of the eigenfrequency of the QNM 1 corresponds to the central frequency of the incident pulse. **(c)** $|\beta_1|$ and $|\beta_2|$ for the on-resonance QNM 1 and off-resonance QNM 2 shown in **(b)**. Circles: reference numerical data obtained directly from Eq. (5). Black solid curve: Results from the EME Eq. (8). Blue dashed and red dashed curves: results respectively obtained by solely considering the CMT-like term (term 1) and the new derivative term (term 2). **(d)** Top panel: $\mathbf{E}_{tot}$ in the resonator, $|x| < 200$ nm, for $t = 65$ fs. Bottom panel: $\mathbf{E}_{tot}$ as a function of time for $x = 150$ nm. Full-wave reference data obtained with COMSOL are shown with black circles and compared with results obtained with the QNM expansion of Eq. (4). The blue-dashed and black-solid curves are respectively obtained with $\beta_m$ values computed with term 1 only and with terms 1 and 2.

Specifically, we examine a 1D Fabry-Perot resonator immersed in air and illuminated by a 10-fs plane-wave Gaussian pulse, as depicted in Fig. 4a. The incident Gaussian pulse is expressed as $\mathbf{E}_b(x,t) = \exp[-i\omega(t + x/c)]\,\mathbf{E}_{env}(x,t)$ with $\mathbf{E}_{env}(x,t) = \exp(-[t + (x - x_d)/c]^2/\tau^2)\,\mathbf{e}_y$ the envelop gaussian term. The central frequency of the pulse $\omega/2\pi = 281$ THz corresponds to the real part of the eigenfrequency of one of the QNM of the Fabry-Perot resonator (QNM 1 in Fig. 4b). Other parameters are $\tau = 2.51$ fs and $x_d = 20$ μm.

Firstly, we compute $|\beta_m|$ with the EME Eq. (8) for two QNMs and compare the results with those obtained by numerically solving Eq. (5). The EME results are shown with the black solid curves in Fig. 4c. They perfectly agree with the results obtained from Eq. (5) that are represented by the black circles. The agreement numerically supports the analytical derivation provided in Section 2.

We further decompose the driving factor $\frac{d}{dt}\langle\mathbf{E}_b|\Delta\boldsymbol{\varepsilon}(x)|\tilde{\mathbf{E}}_m\rangle$ within the EME Eq. (8) into two terms: a CMT-like term proportional to the incident field (term 1) and a new term governed by the time-derivative of the envelope (term 2). Terms 1 and 2 are defined at the top of Fig. 4. As previously discussed, if we neglect the time-derivative term 2 and implement the local excitation approximation, the EME equation reduces to Eq. (10).

To study the importance of the new derivative term, we compute $|\beta_m|$ by separately considering the contributions of terms 1 and 2. When examining the QNM that aligns in frequency with the central frequency of the pulse, denoted as QNM 1, $|\beta_1|$ is accurately determined by solely considering the CMT-like term 1, as shown in the upper panel in Fig. 4c. However, for the off-resonance QNM 2, disregarding term 2 leads to substantial deviation from the exact numerical data for $|\beta_2|$, as depicted in the lower panel of Fig. 4c.

To understand why term 2 exerts minimal influence on resonant QNMs (an observation indirectly encountered in the previous Section), we introduce $\beta'_m$, the envelop of the excitation coefficient, defined by $\beta_m = \beta'_m \exp(-i\tilde{\omega}_m t)$. From the EME Eq. (8), we find that $\beta'_m$ satisfies $\frac{d\beta'_m}{dt} = \exp(i\tilde{\omega}_m t)\left[i\omega\,\langle\mathbf{E}_b|\Delta\boldsymbol{\varepsilon}|\tilde{\mathbf{E}}_m\rangle - \exp(-i\omega t)\left\langle\exp\left(-\frac{i\omega x}{c}\right)\frac{d\mathbf{E}_{env}}{dt}\Big|\Delta\boldsymbol{\varepsilon}\Big|\tilde{\mathbf{E}}_m\right\rangle\right]$. We further ignore the size effect of the Fabry-Perot resonator so that $\mathbf{E}_b(x,t) \approx \mathbf{E}_b(t) = \exp(-i\omega t)\,\mathbf{E}_{env}(t)$ with $\mathbf{E}_{env}(t) = \exp(-[t - x_d/c]^2/\tau^2)\,\mathbf{e}_y$. For the resonant QNM, $\tilde{\omega}_m \approx \omega$, the second term is an anti-symmetric function with respect to $t_d = x_d/c$, resulting in its net contribution to $\beta'_m$ being zero.

Finally, in the two panels of Fig. 4d, we present the reconstructed total field, $\mathbf{E}_{tot}(x,t) = \mathbf{E}_b(x,t) + \mathbf{E}_s(x,t)$, first for a fixed time ($t = 65$ fs) inside the resonator, $|x| < 200$ nm, where the QNM expansion of Eq. (4) is complete [10,11], and then for a fixed coordinate ($x = 150$ nm) as a function of time. The total field reconstructed using 250 QNMs aligns perfectly with reference data obtained with COMSOL time-domain simulations. If we neglect term 2 (blue-dashed curves), a significant deviation compared to COMSOL results is noticeable. This clearly highlights the indispensable role of the new time-derivative term in determining the resonator response.

**Acknowledgements.** PL acknowledge Sander Mann and Andrea Alù for helpful discussions during his visit at CUNY in November 2023. The authors acknowledge Wei Yan for his pivotal role in deriving Eq. (9) when he was postdoc in Bordeaux. They also acknowledge Rachid Zarouf, Thomas Christopoulos, Alejandro Giacomotti and Fabrice Lemoult for helpful feedback during the preparation of the manuscript. PL acknowledges financial supports from the WHEEL (ANR-22CE24-0012-03) Project and the European Research Council Advanced grant (Project UNSEEN No. 101097856).

**Disclosures.** The authors declare no conflicts of interest.

**Data availability.** Data underlying the results presented in this paper are not publicly available at this time but may be obtained from the authors upon reasonable request.